\documentclass[traditabstract]{aa}

\usepackage{amssymb}
\usepackage{color}

\usepackage{txfonts}
\usepackage{graphicx}
\usepackage{natbib}
\bibpunct{(}{)}{;}{a}{}{,}

\newcommand{\ratio}{$\dot{N}_{LT}$/$\dot{N}_{FP}$}

\title{Implications for electron acceleration and transport from
  non-thermal electron rates at looptop and footpoint sources in
  solar flares}

\author{P.~J.~A.~Sim\~oes \and E.~P.~Kontar}

\institute{SUPA, School of Physics and Astronomy, University of Glasgow, G12 8QQ, UK}

\date{Received / Accepted}

\abstract{The interrelation of hard X-ray (HXR) emitting sources and
  the underlying physics of electron acceleration and transport
  presents one of the major questions in high-energy solar flare
  physics. Spatially resolved observations of solar flares often
  demonstrate the presence of well-separated sources of bremsstrahlung
  emission, so-called coronal and footpoint sources. Using spatially
  resolved X-ray observations by the Reuven Ramaty High Energy Solar
  Spectroscopic Imager (RHESSI) and recently improved imaging
  techniques, we investigate in detail the spatially resolved electron
  distributions in a few well-observed solar flares. The selected
  flares can be interpreted as having a standard geometry with
  chromospheric HXR footpoint sources related to thick-target X-ray
  emission and the coronal sources characterised by a combination of
  thermal and thin-target bremsstrahlung.  Using imaging spectroscopy
  techniques, we deduce the characteristic electron rates and spectral
  indices required to explain the coronal and footpoint X-ray
  sources. We found that, during the impulsive phase, the electron
  rate at the looptop is several times (a factor of $1.7-8$) higher
  than at the footpoints. The results suggest that a sufficient number of
  electrons accelerated in the looptop explain the precipitation
  into the footpoints and imply that electrons accumulate in the
  looptop. We discuss these results in terms of magnetic trapping,
  pitch-angle scattering, and injection properties. Our conclusion is
  that the accelerated electrons must be subject to magnetic trapping
  and/or pitch-angle scattering, keeping a fraction of the population
  trapped inside the coronal loops. These findings put strong
  constraints on the particle transport in the coronal
  source and provide quantitative limits on deka-keV electron
  trapping/scattering in the coronal source.
}

\titlerunning{Electron rate at looptop and footpoint sources of solar flares}
\authorrunning{Sim\~oes \& Kontar}

\begin{document}
\maketitle

\section{Introduction}
\label{sec:intro}

The X-ray images of solar flares with spectral capabilities taken by
the Hard X-ray Telescope (HXT) on board of Yohkoh
\citep{1991SoPh..136...17K} and Reuven Ramaty High Energy Solar
  Spectroscopic Imager (RHESSI) \citep{2002SoPh..210....3L}
have revealed a wealth of information on spatially resolved
distributions of energetic electrons in solar flares \citep[see
  e.g.][as recent
  reviews.]{2011SSRv..159....3D,2011SSRv..159..107H,2011SSRv..159..301K}
These spatially resolved observations indicate appearances of distinct
coronal and footpoint X-ray sources. Hard X-ray (HXR) footpoint
sources are usually described in terms of collisional thick-target
models
\citep{1963SPhD....8..543D,1969sfsr.conf..356E,1968ApJ...151..711A,1969SoPh....8..341L,1971SoPh...18..489B,1972SvA....16..273S,1973SoPh...31..143B,1976SoPh...50..153L},
where the electrons accelerated in the corona travel downwards inside
magnetic loops marking footpoint sources in dense chromospheric
plasma by emission in X-rays, EUV, and white-light emissions
\citep{2008uxss.book.....P}.  The observations show that the HXR source
height in the chromosphere decreases with increasing X-ray energy
\citep{1992PASJ...44L..89M,2002SoPh..210..383A,2008A&A...489L..57K,2010ApJ...721.1933S,2011ApJ...735...42B}
as anticipated from downward propagating electron beams.  In addition,
the characteristic sizes of HXR sources decrease with increasing
energy, which is consistent with collisional electron transport along
converging magnetic flux tubes
\citep{2008A&A...489L..57K,2011ApJ...735...42B,2011ApJ...740L..46F,2012ApJ...752....4B,2012ApJ...750L...7X}.
Coronal X-ray sources are prominently visible at and below $\sim
10-30$ keV and are usually due to thermal and non-thermal
bremsstrahlung, which indicates plasma heating and the presence of
non-thermal electron populations \citep[e.g.][]{2003ApJ...595L.107E}.
Spectroscopically, there is no clear boundary between the two
processes, with the spectrum transiting smoothly from thin-target
emission in the corona to thick-target emission within the range $10-30$
keV. The looptop sources are
believed to be connected to the process of energy release in flares
\citep[e.~g.][]{2002A&ARv..10..313P,2011SSRv..159..107H}.  Some models
place the magnetic reconnection at the looptop, where the X-ray
emission is a consequence of the plasma heating and particle
acceleration
\citep{1998A&A...334.1112J,1999ApJ...527..945P,2007ApJ...666.1256B,2011ApJ...730L..22K,2012A&A...543A..53G}.
Other models propose that the X-point is placed higher in the
corona, and the looptop X-ray source shows that the interaction between a
downward reconnection flow and plasma fills a soft X-ray loop
\citep{1997ApJ...478..787T}.  The recent observations of
energy-dependent HXR source sizes inside the dense coronal loops
\citep{2008ApJ...673..576X,2011ApJ...730L..22K,2011A&A...535A..18B,2012A&A...543A..53G,2012ApJ...751..129T}
also suggest within-the-loop electron acceleration and transport.

\citet{2002ApJ...569..459P} presented a comprehensive analysis of 18
Yohkoh events, finding that the spectral index of the looptop source
is softer than the footpoints on the average by about 1, although this
result is limited by the poor energy resolution of the Yohkoh images,
with only four energy ranges \citep{1991SoPh..136...17K}.  With the
launch of RHESSI, detailed imaging spectroscopy with energy resolution
$\sim 1$~keV has become widely available
\citep[e.g.][]{2003ApJ...595L.107E,2006A&A...456..751B,2007ApJ...665..846P,2009SoPh..260..135K}.
Thus, \citet{2003ApJ...595L.107E} found that for the GOES X-class 2002
July 23 flare the coronal source is consistent with the thermal source
of around $4\times 10^{7}$~K and the footpoint-like sources that have
power-law spectra with different spectral
indices. \citet{2006A&A...456..751B} presented a systematic study of
the relation between looptop and footpoint sources for a few flares
observed by RHESSI.  They found soft-hard-soft behaviour in both the
coronal source and the footpoints. Assuming the scenario where
  the same electron population produces the coronal source by
  thin-target bremsstrahlung and the chromospheric source by
  thick-target bremsstrahlung, one would expect a difference of
  $\simeq 2$ between the two photon spectral indices. They found that
the average of all mean differences is $\simeq 1.8$. However, the
difference is considerably larger than 2 for two out of five events,
and smaller than 2 for the other three events analyzed. The coronal
source is nearly always softer than the footpoints. The footpoint
spectra differ significantly only in one event out of five. While the
observations are consistent with acceleration in the coronal source
and a subsequent propagation into the chromosphere, their result
excludes the simplest scenario of electrons free-streaming from the
acceleration region towards the footpoints, which indicates that other
transport processes must be considered, e.~g. wave-particle
interaction can account for spectral differences larger than 2
\citep{2011A&A...529A.109H}.  \citet{2007A&A...461..315T} used
HXT/Yohkoh data of 117 flares to investigate the ratio of photon
counts from looptop to footpoint sources, finding a correlation
between this ratio and height and attributing the effect of loop
convergence in 80\% of the flare sample to inferred mirror ratios
lower than 2.1.  Coronal trapping is a natural consequence of the
expected convergence of the magnetic field between the corona and the
chromosphere. Trap-plus-precipitation models have been proposed and
developed
\citep{1966PASJ...18...57T,1976MNRAS.176...15M,1981ApJ...251..781L,1983ApJ...269..715L,1988A&A...194..279M,1990A&A...235..431A,1990A&A...234..487M,1991A&A...242..256M,1997A&A...326.1259F,1998ApJ...505..418F,2010SoPh..266..323P}. 
Also, considerable observational evidence of trapping has been
provided
\citep{1994ApJ...436..941R,1997ApJ...489..442A,1998ApJ...502..468A,1999ApJ...522.1108M,2000ApJ...531.1109L,1999ApJ...517..977A,2000ApJ...543..457L,2002ApJ...572..609L,2002SoPh..210..323A,2002ApJ...580L.185M,2011ApJ...731L..19F,2011ApJ...740...46H}.
Analysing microwave brightness maps, \citet{2002ApJ...580L.185M}
applied a collisionless adiabatic trap model to explain looptop
sources at 17 and 34 GHz observed in four flares. They concluded that
the electron distribution along the loops must be highly
inhomogeneous, with the energetic electrons concentrated at the upper
part of the loop. On the other hand, microwave loop-like brightness
structures of different flares could only be reconciled with models of
uniform magnetic field and homogeneous spatial distribution of
non-thermal electrons
\citep{2001ApJ...557..880K}. \citet{2000ApJ...533.1053N} successfully
reproduced the main spatial morphology of 5 and 15 GHz brightness maps
of a weak flare using a non-uniform magnetic field model and uniform
distribution of electrons, in agreement with model simulations
\citep{1984A&A...139..507A,1984A&A...141...67K,2006A&A...453..729S,2010SoPh..266..109S}. Those
examples evidence how trapping conditions can change from flare to
flare, influencing the electron distribution along flaring
loops. Moreover, it seems that the microwave emission maps indirectly
show the trapped population, as the emission also depends strongly on
the magnetic strength. However, HXR images will mainly show the
precipitated population as footpoint sources, only eventually showing
a non-thermal HXR coronal source. As the collisions are
energy-dependent, one can expect that deka-kev HXR-producing electrons
and MeV microwave-producing electrons will evolve in slightly
different ways during a solar flare
\citep[e.g.][]{2002ApJ...572..609L,2012SoPh..tmp..296G}.

Importantly, electron scattering (often strong) is required to achieve
efficient acceleration of particles
\citep[e.g.][]{1997JGR...10214631M,1998SSRv...86...79M,1999ptep.proc..135P,2012ApJ...754..103B}
and to explain the lack of strong HXR anisotropy in observations,
while the transport of particles affects how the particles are
accelerated \citep[see
  e.g.][]{2009ApJ...692L..45B,2012ApJ...754..103B,2012SSRv..tmp...49P}.

Despite the importance of comparative studies between the coronal
sources and footpoints, there were no systematic comparisons of the
number of electrons in the footpoints and the coronal sources for the
events with visible footpoints. We emphasize that the inference of
imaging spectroscopy parameters relies strongly on the precise
knowledge of the X-ray source area, which was only developed fairly
recently
\citep[e.g.][]{2007SoPh..240..241S,2008A&A...489L..57K,2009ApJ...698.2131D,2010ApJ...717..250K}.
In addition, the full Spectral Response Matrix (SRM) of RHESSI detectors
(including non-diagonal terms) for imaging spectroscopy became
available relatively recently, since 2006 February.

We present a comparative analysis of the electron distributions in
footpoints and coronal sources for four well-observed flares.  We
find for the first time the electron rate required to explain the
coronal and chromospheric X-ray emissions.  In Sect. \ref{sec:data} we
present the observations and in Sect. \ref{sec:method} the methodology
to retrieve the electron rates from imaging spectroscopy. Our results
are presented in Sect. \ref{sec:results} and discussed in
Sect. \ref{sec:discussion}. The work is summarized in
Sect. \ref{sec:conclusion}.

\section{Observational data}
\label{sec:data}

We selected four well-observed events in which the common loop
structure could be identified: two footpoint sources at higher photon
energies and a looptop source at lower energies. We avoided events
where looptops and footpoints overlap due to the projection of the
loop geometry. Also, we only selected flares where a non-thermal
source at the looptop could be resolved. The flare list
with details is presented in Table \ref{tab:eventinfo}. Different
aspects of the selected flares were previously investigated by other
authors: 2002 July 23 \citep{2003ApJ...595L.107E}, 2003 November 02
\citep{2007SoPh..245..311S}, 2011 February 24
\citep{2011A&A...533L...2B}.

\begin{table*}
  \caption{Selected solar flares: date and time used for analysis, GOES
    peak and classification events, RHESSI spectral fitting results,
    source sizes measured from HXR images, and electron rate results
    for looptop $\dot{N}_{LT}$ and footpoint $\dot{N}_{FP}$
    sources.}
  \label{tab:eventinfo}
  \centering
  \begin{tabular}{llcccc}
    \hline\hline
    Flare & &A & B & C & D\\
    \hline
    Date & &2002 July 23  & 2003 Nov 02 & 2011 Feb 24  & 2011 Sept 24 \\
    Time &UT &00:27:26 & 17:15:54 & 07:29:40 & 09:35:53 \\
    $\Delta t$& sec & 284 & 246 & 176 & 82\\
    GOES peak && X5.1 & X8.3 & M3.5 & X1.9\\
    GOES time &UT& 00:36 & 17:24 & 07:35 & 09:40 \\
    CLEAN beam && 2.3 & 2.0 & 1.9 & 2.4\\
    \hline
    Imaging spectroscopy fitting results & & & & & \\
    $<\bar n VF_0>$&$10^{55}$cm$^{-2}$s$^{-1}$ &$77 \pm 12$ & $169 \pm 12$ & $0.7\pm 0.1$ & $7 \pm 2$ \\
    $\delta_{LT}$  & &$3.7  \pm 0.2$  & $2.9 \pm 0.1$   & $3.3\pm 0.2$ & $3.6 \pm 0.4$\\
    $\dot{N}_{FP} $ & $10^{35}$s$^{-1}$ & $32 \pm 5$ & $153 \pm 11$ & $1.1 \pm 0.1$ & $ 5.2 \pm 1.1$\\
    $\delta_{FP}$    & &$4.0\pm 0.1$   & $3.9\pm 0.1$   & $3.8\pm 0.1$ & $3.8 \pm 0.1$\\
    \hline
    Spatially integrated spectral fitting results & & & & & \\
    $EM$ &$10^{49}$cm$^{-3}$& $3.8\pm 0.4$ &$4.2\pm 0.6$ & $0.24\pm 0.02$ & $0.8 \pm 0.1$\\
    $T$  &MK& $33$ & $34$ &$21$ & $34$\\
    $\dot{N}$ & $10^{35}$s$^{-1}$&$27\pm 3$ & $122\pm 11$ &$2.0\pm 0.1$ &$8.9 \pm 1.0$\\
    $\delta$ & &$3.9 \pm 0.1$& $4.0\pm 0.1$ & $4.0\pm 0.1$ & $3.8 \pm 0.1$\\
    $\bar n$ & $10^{11}$cm$^{-3}$ & $2.7$ & $1.5$ & $0.6$ & $1.9$ \\
    \hline
    Characteristic sizes of the sources& & & & & \\
    $A$ &  $10^{18}$cm$^2$ & 0.6 &  2.2 & 0.9 & 0.3 \\
    $D$ & $10^{8}$cm &8.9  & 8.0 & 8.3 & 6.4 \\
    $V_{th}$ & $10^{27}$cm$^3$  & 0.5 & 1.8 & 0.8 & 0.2\\
    $L$      & $10^8$ cm   & 5.2 &  4.3 & 8.3& 4.7\\
    \hline
    Electron rates& & & & & \\
    $\dot{N}_{LT}$&$10^{35}$s$^{-1}$  & $54 \pm 11$ & $258 \pm 38$ & $1.6 \pm 0.3$ & $8 \pm 3$ \\
    $\dot{N}_{FP}$ (fully ionized FP, without albedo) &$10^{35}$s$^{-1}$  & $32 \pm 5$ & $153 \pm 11$ & $1.1 \pm 0.1$ & $ 5.2 \pm 1.1$\\
    \ratio \ (as above) & & $1.7 \pm 0.6$ & $1.7 \pm 0.4$ & $1.5 \pm 0.4$ & $1.4 \pm 0.9$\\


    $\dot{N}_{FP}$ (neutral FP, with albedo) &$10^{35}$s$^{-1}$ & $6\pm 1$& $27\pm 2$&$0.4\pm 0.1$& $0.8\pm 0.2$\\
    \ratio \ (as above) &&$9.3 \pm 3.0$&$10.1 \pm 2.3$&$4.5\pm 1.4$& $7.8 \pm 5.2$\\
    \hline
  \end{tabular}
\end{table*}

\section{Methodology}
\label{sec:method}

\subsection{Imaging spectroscopy}
\label{sec:imgsp}
\begin{figure*}
\resizebox{\hsize}{!}{\includegraphics[angle=0]{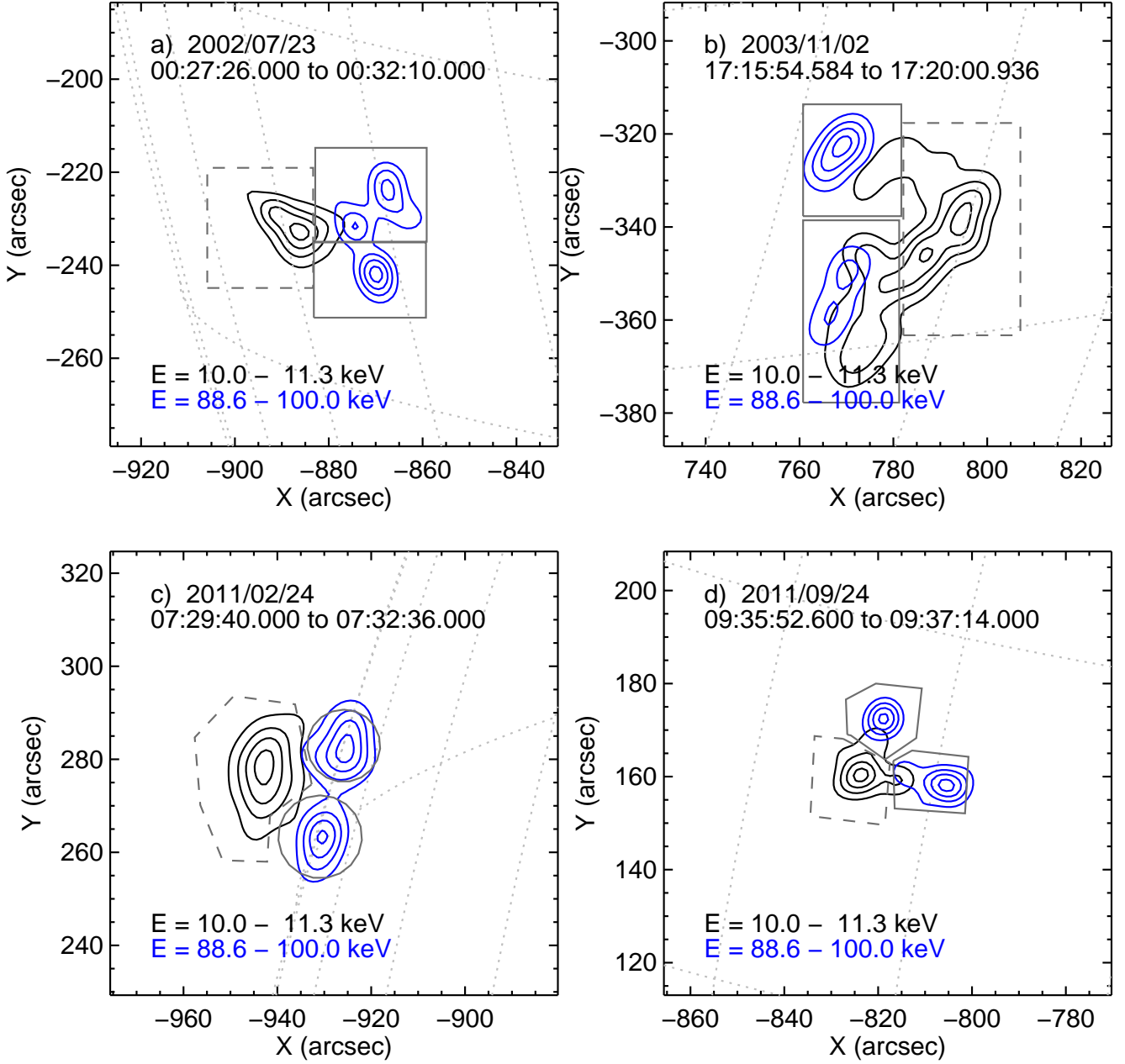}}
\caption{CLEAN maps for each flare: ({\em a}) 2002 July 23, ({\em b})
  2003 November 02, ({\em c}) 2011 February 24, and ({\em d}) 2011
  September 24, showing the emission at 10-11.3 keV (black contours)
  and 88.6-100keV (blue contours) at 30, 50, 70, 90\% of maximum of
  each energy bin. The ROIs for the imaging
  spectroscopy are defined in gray lines: footpoints (continuous
  lines) and looptop (dashed lines). Integration times are shown in
  each frame and in Table \ref{tab:eventinfo}. The CLEAN beam width factor
  used to construct each map according to the best
  agreement with visibility forward-fitting is also presented in Table
  \ref{tab:eventinfo}. For each, a total of 19 maps were made with
  energy bins logarithmically spaced between 10 and 100 keV.}
\label{fig:mapsA}
\end{figure*}

For each of the four flares, we constructed CLEAN images
\citep{2002SoPh..210...61H} with a pixel size of $1''$, using front
detectors 3 to 8, for 19 logarithmically binned energy bands from 10
to 100 keV. Since we rely on the source sizes taken from the
reconstructed images, we verified the best CLEAN beam width for each
flare. \cite{2009ApJ...698.2131D} and \cite{2010ApJ...717..250K} have
pointed out that CLEAN images usually have systematically larger sizes
than other algorithms when using the default beam width factor of 1.0.
To ensure the best possible determination of the source sizes using
CLEAN images, we applied the visibility forward-fitting procedure
\citep{2007SoPh..240..241S} on the footpoints of each flare, adjusted
the CLEAN beam size, and re-calculated the images until the FWHM of
the footpoint from both algorithms had the same size. The integration
time interval during the impulsive phase and the CLEAN beam used for
each event are presented in Table \ref{tab:eventinfo}. The CLEAN beam
factor found for these flares lay in the range $1.9-2.4$, suggesting
the optimum value is $\simeq 2$
\citep{2009ApJ...698.2131D,2010ApJ...717..250K}, instead of the
default value of 1. Fig. \ref{fig:mapsA} shows CLEAN maps for the four
flares at lowest (10-11.3 keV) and highest (88.6-100 keV) energy bins.

\subsubsection{Regions of interest selection}
Within each image, we selected regions of interest (ROI) which capture
looptop (LT) and footpoint (FP) sources for further imaging
spectroscopy using OSPEX \citep{2002SoPh..210..165S}.  For the flares
considered here, two footpoints can be easily identified and noted in
the literature, e.~g. \cite{2003ApJ...595L.107E},
\cite{2007SoPh..245..311S}, and \cite{2011A&A...533L...2B} for flares
A, B, and C, respectively. The looptop ROI was chosen to include the main
emitting regions at low energies, but without being too large, to
avoid contribution from the FPs. The looptop and footpoint ROIs for each flare
are shown in Fig. \ref{fig:mapsA}. The background was estimated taking
the sum of pixels outside the ROIs in each image, then normalized for
the area of each ROI, similar to an approach applied by
\cite{2006A&A...456..751B}. This method probably overestimates the
background noise, especially at lower energies; nevertheless, it is
useful to identify the reliable energy ranges for the fitting
procedure.

\subsubsection{Spectral fitting}
While the detailed analysis of HXR spectrum in some events, e.~g. 2002
July 23 flare, demonstrated that the actual spectrum does contain
deviations from a simple power-law
\citep{2003ApJ...595L.127P,2011SSRv..159..301K}, we assume a single
power-law in the electron spectrum for the purposes of our analysis.
For each flare, the looptop source was fitted with a thin-target model and
the footpoint sources were fitted using a thick-target model \citep[see][for
  details]{2011SSRv..159..301K}.  Assuming looptop source is a thin-target
emission, the photon flux detected at the Earth is the convolution of
the electron distribution and the bremsstrahlung cross-section.

Following \cite{1971SoPh...18..489B} and \cite{2003ApJ...595L.115B}, we define
the mean target proton density $\bar n = V^{-1}\int n({\bf r}) \, dV$,
where $n({\bf r})$ is the plasma density, $V$ is the volume of the LT
emitting region. We can then write the photon spectrum at distance the Earth
as
\begin{equation}
  I_{LT}(\epsilon) = {1 \over 4\pi R^2} \, \,
  \int_\epsilon^\infty  <\bar n V \bar F(E)> \, Q(\epsilon,E)\, dE,
\label{def_LT}
\end{equation}
where the bremsstrahlung cross-section $Q(\epsilon, E)$ following
\citep{1997A&A...326..417H} and ${\bar F}(E)$ is the {\it mean
  electron flux spectrum} defined by
\begin{equation}
\bar F(E)= \frac{1}{\bar n V} \int_V F(E,{\bf r}) \, n({\bf r}) \, dV.
\end{equation}
  From the OSPEX \citep{2002SoPh..210..165S} fit of the photon flux
  from looptop ROI with the model given by Eq. (\ref{def_LT}), one finds
  $<\bar n V\bar{F}(E)>$ without any additional assumptions, where
\begin{equation}
<\bar n V\bar{F}(E)>=<\bar n V\bar{F}_0>\frac{\delta
  _{LT}-1}{E_0}\left(\frac{E}{E_0}\right)^{-\delta _{LT}}, \;\;
E\geqslant E_0
\label{eq:nVF_def}
\end{equation}
normalised to $<\bar n V\bar{F}_0>$ [electrons~cm$^{-2}$~s$^{-1}$],
which is the product of the mean target density $\bar n$ and the
energy-integrated (above the low energy cut-off $E_0$) mean electron
flux $\bar F_0 = \int \bar F(E) dE$ in the volume $V$. The energy-integrated electron rate
$\dot{N}_{LT}$ [electrons~s$^{-1}$] required to explain the observed
thin target emission in the looptop is simply
\begin{equation}
\dot{N}_{LT}=\bar F_0 S,
\label{eq:nlt1}
\end{equation}
where $S$ is the cross-section area of the loop. We can estimate $\bar F_0$ by
\begin{equation}
 \bar F_0 = \frac{<\bar n V \bar F_0>}{\bar n V}= \frac{<\bar n V \bar F_0>}{\bar n LS},
\label{eq:f0}
\end{equation}
where $L$ is the length of the looptop source in the direction along the
loop (hence $V=LS$). Substituting Eq. \ref{eq:f0} into Eq. \ref{eq:nlt1} we have
\begin{equation}
  \dot N_{LT} = \frac{<\bar n V \bar F_0>}{\bar n L}.
\label{eq:nlt}
\end{equation}
The thermal plasma density $\bar n$ and the size
$L$ are obtained from observation data (see Sect. \ref{sec:images}).

\begin{figure*}
\resizebox{\hsize}{!}{\includegraphics[angle=0]{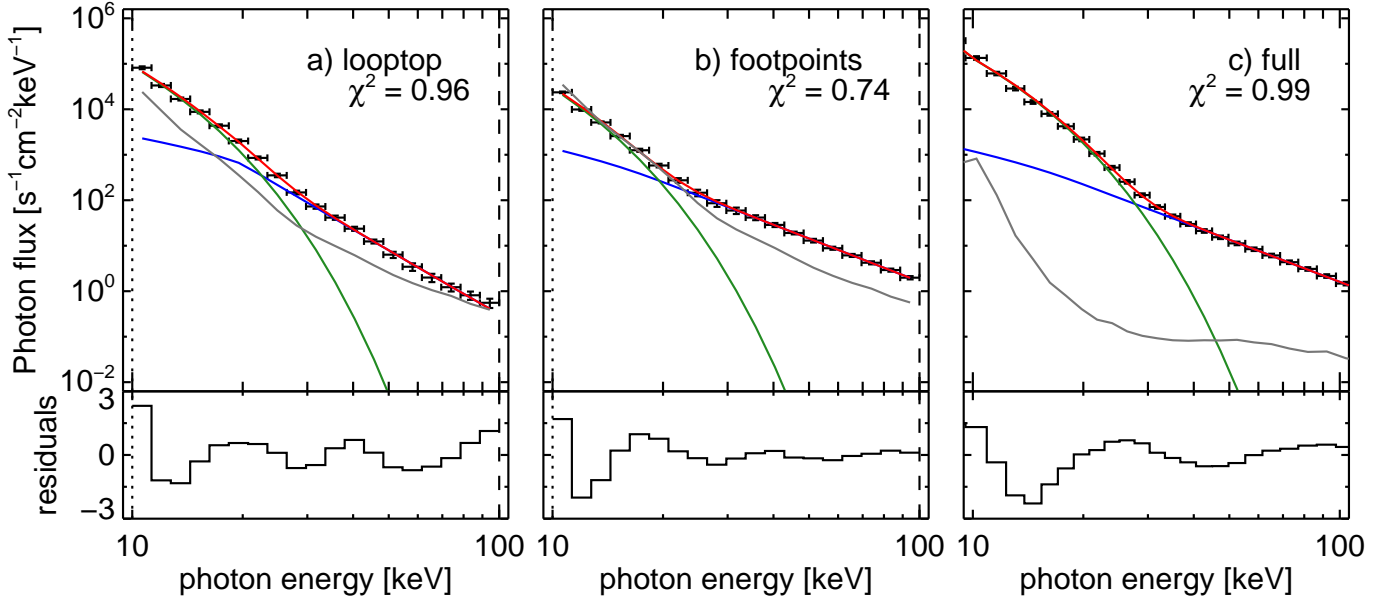}}
\caption{looptop ({\em a}), footpoint ({\em b}) sources and
  spatially integrated ({\em c}) HXR spectra (black) of flare
  A, 2002 July 23, estimated background (gray), and fitting result
  (red) and its components: thermal (green) and non-thermal
  (blue). For the looptop spectrum, the non-thermal is a thin-target
  model; for the footpoint and integrated spectra, it is a
  thick-target model.}
\label{fig:imgspecA}
\end{figure*}

\begin{figure*}
\resizebox{\hsize}{!}{\includegraphics[angle=0]{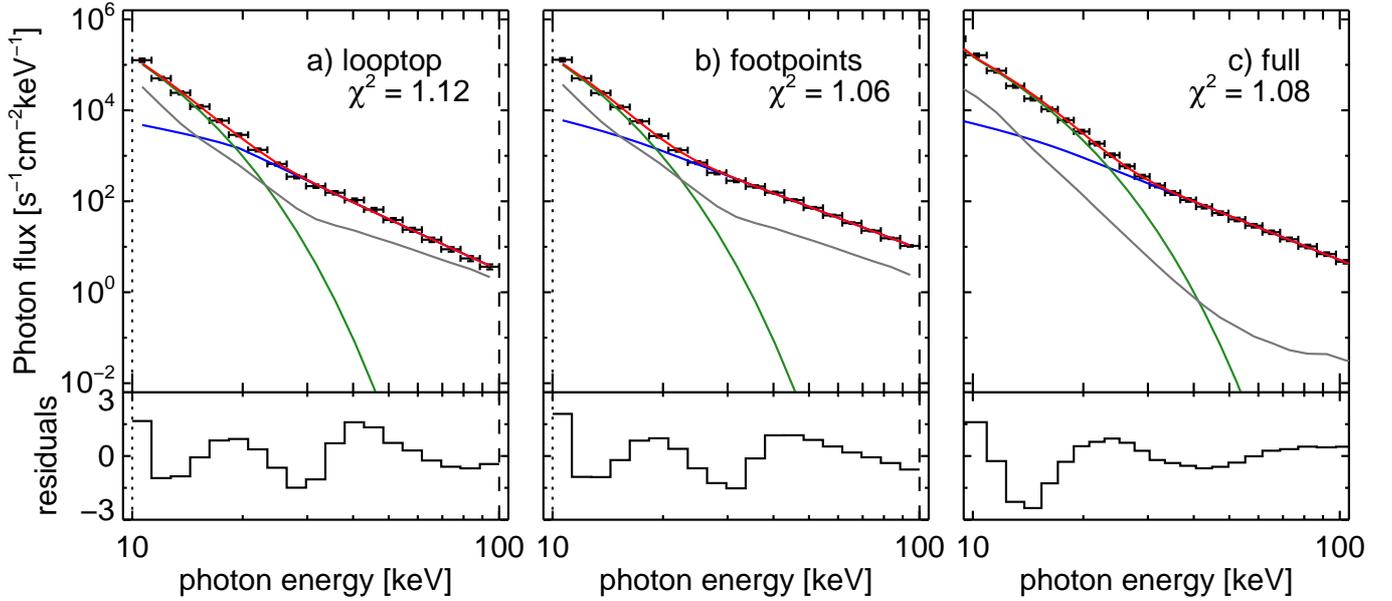}}
\caption{Same as Fig. \ref{fig:imgspecA}, for flare B, 2003 November 02.}
\label{fig:imgspecB}
\end{figure*}

\begin{figure*}
\resizebox{\hsize}{!}{\includegraphics[angle=0]{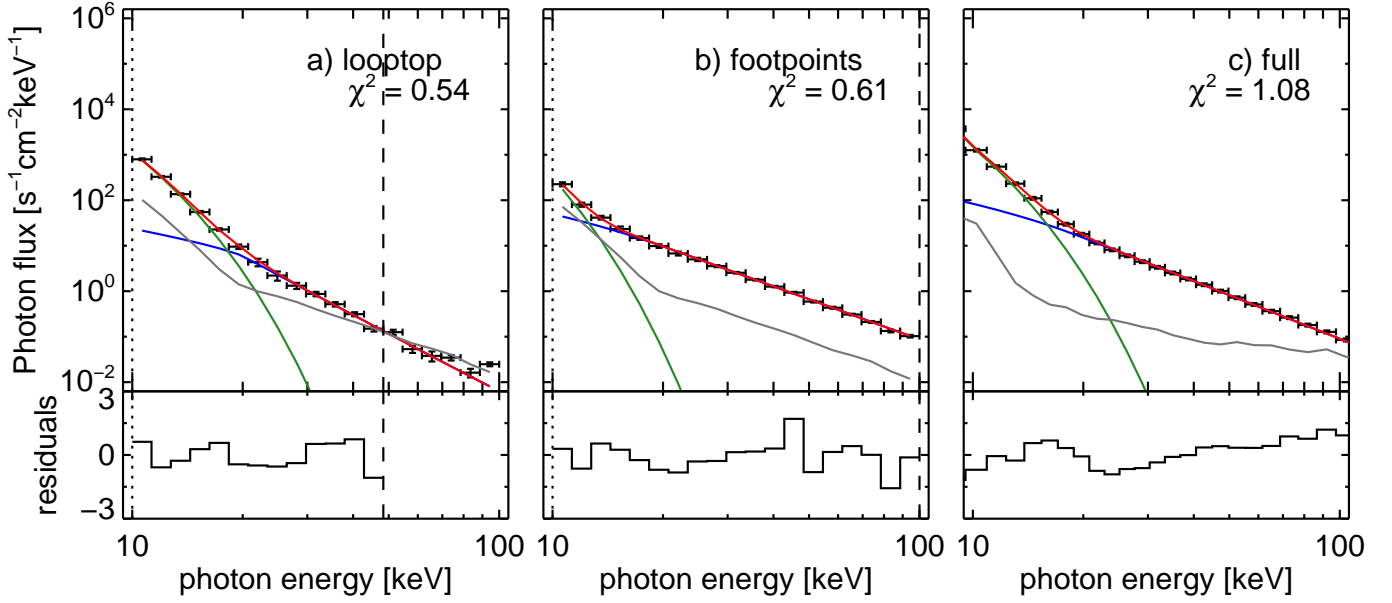}}
\caption{Same as Fig. \ref{fig:imgspecA}, for flare C, 2011 February 24.}
\label{fig:imgspecC}
\end{figure*}

\begin{figure*}
\resizebox{\hsize}{!}{\includegraphics[angle=0]{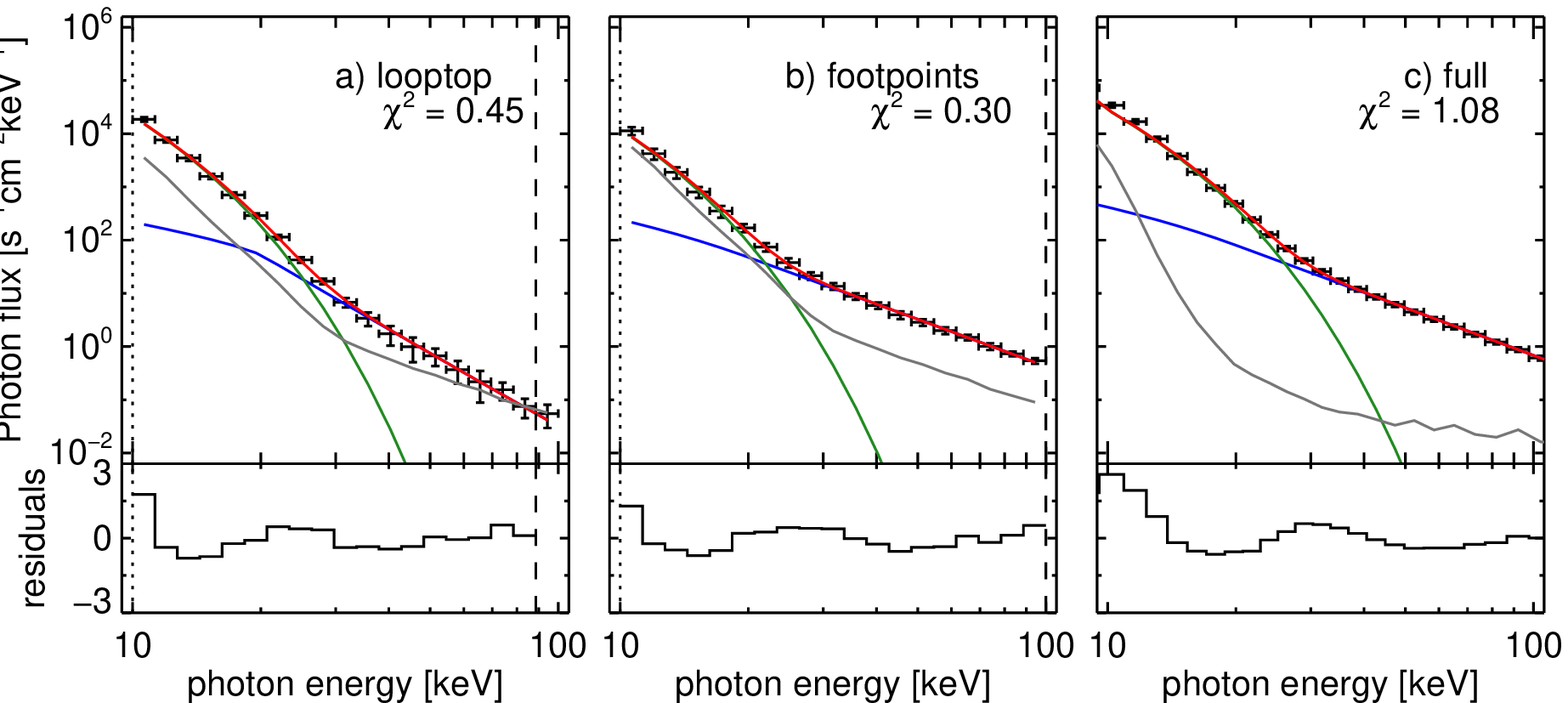}}
\caption{Same as Fig. \ref{fig:imgspecA}, for flare D, 2011 September 24.}
\label{fig:imgspecD}
\end{figure*}

For the footpoint emission, we assume a thick-target model, so that
the observed photon flux at the distance $R$ is given:
\begin{equation}
I_{FP}(\epsilon) = {1 \over 4\pi R^2} \, \, \int_\epsilon^\infty
\dot{N}_{FP} {E_0 \over K}\left(\frac{E}{E_0}\right)^{\delta _{FP}-2}
\, Q(\epsilon,E)\, dE,
\label{def_FP}
\end{equation}
where $\dot{N}_{FP}$ is the electron precipitation rate at footpoints
[electrons~s$^{-1}$], and $K=2\pi e^4 \Lambda$, $e$ is the electronic
charge and $\Lambda$ is the Coulomb logarithm (assumed $\simeq 22.7$
for fully ionized chromosphere and $7.1$ for neutral chromosphere, see
Sect. \ref{sec:albedo}). Here we emphasize that $\delta_{FP}$ and
$\dot{N}_{FP}$ are the spectral index and the electron precipitation rate
of the electrons entering into the footpoints. If the electrons are
unaffected by the transport within the loop, this spectral index
should match $\delta _{LT}$ and the electron rate that of looptop source
$\dot{N}_{LT}$.  Both fits included a thermal component, using
full CHIANTI model \citep{2009A&A...498..915D}, with continuum and
spectral lines. A spatially integrated spectrum was also made for each
flare using the same time intervals and fitted with a thermal plus
thick-target component. The thick-target model was set as a
single power-law. The fitting results (Eqs. \ref{def_LT} and
\ref{def_FP}) for the imaging and spatially integrated spectroscopy
are presented in Table \ref{tab:eventinfo}. The photon spectra of the
LT, FP, and spatially integrated sources are plotted in
Figs. \ref{fig:imgspecA} to \ref{fig:imgspecD} for each flare,
respectively, showing the fitting components (thermal and
non-thermal) and fitting residuals.


\subsubsection{Image-derived parameters}
\label{sec:images}
The overall geometry of the selected flares can be interpreted as a
single loop-like magnetic structure with two footpoints and a coronal
source at the apex of the loop. We defined a loop baseline as the line
connecting the two footpoints, which is used as a reference to measure
the sizes of the sources at the looptop. The thermal plasma density
can be calculated using the observations of thermal X-ray
emission. The isothermal fit to the spatially integrated RHESSI
spectrum (right-hand frames in Figs. \ref{fig:imgspecA} to
\ref{fig:imgspecD}) gives emission measure $EM=\bar n^2V_{th}$, where
the volume of the thermal source can be estimated as
$V_{th}=AD$, $A$ is the area of the source, $D$ is the
cross-section diameter of the source, and assuming uniform plasma
density $\bar n$ within the volume $V_{th}$. Using the emission maps
at 10-11.3 keV, $D$ was estimated by measuring the length of this
source at 50\% level in the direction orthogonal to the baseline and
$A$ as the source area at 50\% level of the maximum emission. Using
those measurements, the plasma density is then obtained by $\bar
n=(EM/DA)^{1/2}$ and directly inserted into Eq. \ref{eq:nlt},
resulting in
\begin{equation}
\dot{N}_{LT}=\left(\frac{DA}{EM}\right)^{1/2}\frac{<\bar nV\bar{F}_0>}{L}.
\label{eq:nlt2}
\end{equation}
Similar results for $EM$ are obtained using the imaging
spectroscopy and selecting looptop regions only (left-hand frames of
Figs. \ref{fig:imgspecA} to \ref{fig:imgspecD}), but the imaging
spectroscopy gives larger uncertainties.  The values found for the
plasma density $\bar n/ 10^{11}$ cm$^{-3}$ are $2.7$, $1.5$,
$0.6$, and $1.9$ for flares A, B, C, and D,
respectively. The geometry aspects of the flares and plasma density
values inferred are presented in Table \ref{tab:eventinfo}.

\begin{figure*}
\resizebox{\hsize}{!}{\includegraphics[angle=0]{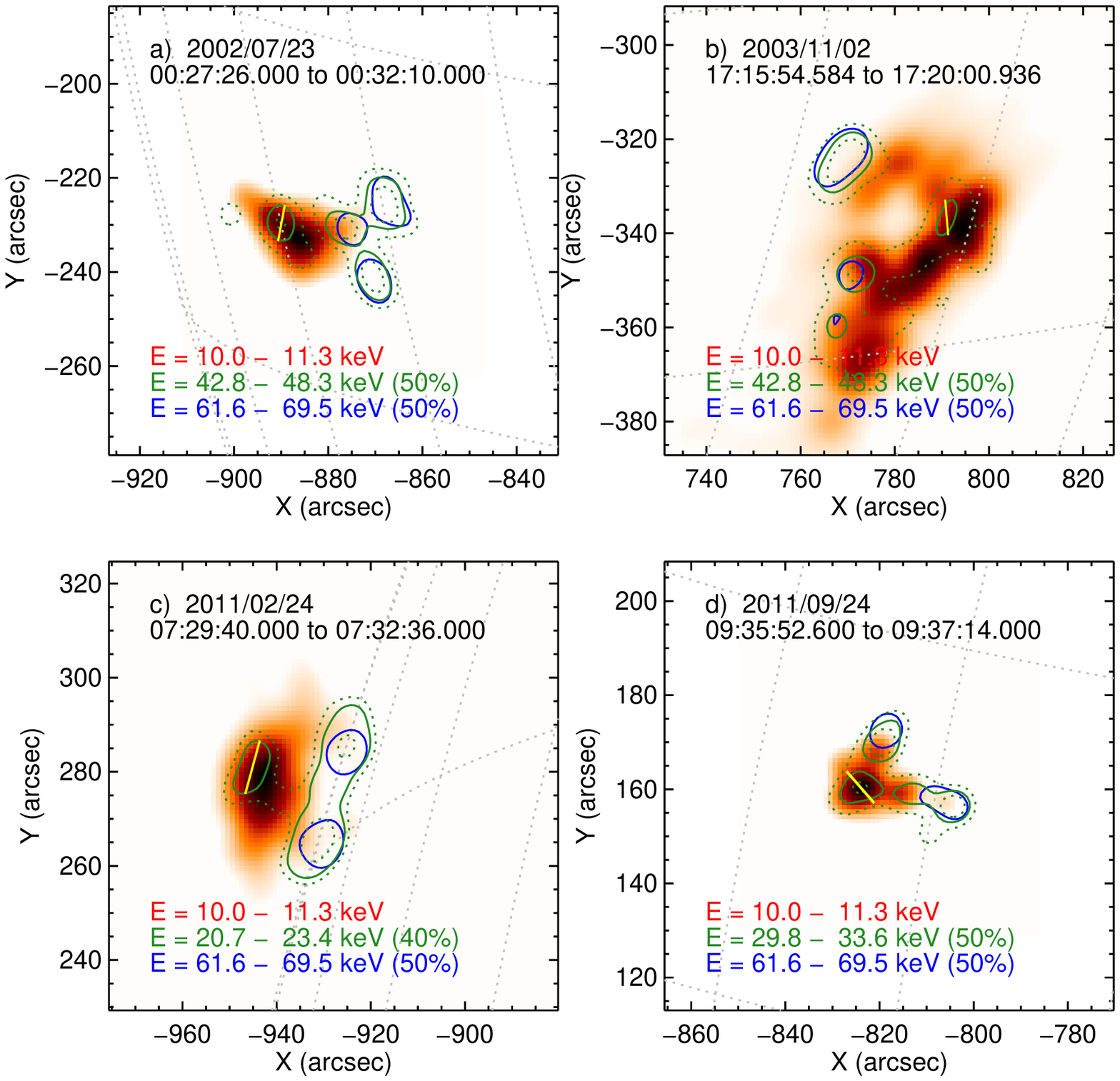}}
\caption{CLEAN maps for the four flares, evidencing thermal
  emission at 10-11.3 keV ({\em red}), footpoints at 61.6-69.5
  keV ({\em blue} contours, at 50\% of the maximum of each map), and
  non-thermal looptop sources ({\em green solid}) at 42.8-48.3 keV
  (50\%) for flares A and B, 20.7-23.4 keV (40\%) for flare C, and
  29.8-33.6 keV (50\%) for flare D. The countour levels of 30\% and 70\%
  at those energies (green dashed lines) are also shown. The thick
  yellow line indicates the length $L$. Date and integration times are
  indicated in each map.}
\label{fig:resA}
\end{figure*}

We define the length $L$ of the non-thermal source at the looptop as
the size (at 50\% of maximum emission) in the direction parallel to
the footpoint baseline and crossing the point with the maximum
emission. $L$ is measured at a sufficiently high energy, where the
contribution from thermal emission is not significant and where the
source is spatially resolved. The non-thermal looptop sources can be
easily identified at $42.8-48.3$ keV, $42.8-48.3$ keV, $20.7-23.4$
keV, and $29.8-33.6$ keV for each flare, respectively
(Fig. \ref{fig:resA}); the measured lengths $L$ for each flare are
presented in Table \ref{tab:eventinfo}.

\section{Observational results and analysis}
\label{sec:results}

Using Eq. \ref{eq:nlt2} with the values found above,
the electron rate at the looptop $\dot{N}_{LT}$ can be calculated and
is summarized in Table \ref{tab:eventinfo}, along with
the electron rate at the footpoints obtained using thick-target forward
fit for the photon flux spectrum.

Spectral index analysis of these four flares reveals that all flares
have a similar footpoint spectral index, which is close to 3.9 (see
Table \ref{tab:eventinfo}). The differences between the spectral index
are $\delta_{LT}-\delta_{FP}=$~$-0.3 \pm 0.2 $,~$-1.0 \pm 0.1$,~$-0.5 \pm
0.2$, and~$-0.2 \pm 0.4$ for each flare. This appears consistent with previous
findings, e.g. \citet{2003ApJ...595L.107E} reported that the photon
spectral index difference is less than 2 for the 2002 July 23 flare and
\citet{2006A&A...456..751B} reports a fairly broad range of spectral
index differences between footpoints and coronal sources.

In general, our results suggest that in all flares the precipitation rate
into footpoints is smaller than the electron rate required to explain the
coronal HXR emission,
i.e. $\dot{N}_{LT}/\dot{N}_{FP}>1$. Assuming fully ionized
chromospheric footpoints and no albedo, $\dot{N}_{LT}/\dot{N}_{FP}$
for the four flares appears around 1.6, which is slightly larger than
unity. This is expected in an idealized model, where non-thermal electrons
propagate without interaction, so that $\dot{N}_{LT}\simeq
\dot{N}_{FP}$. However, as discussed in the next section, the effects
of neutral chromosphere and albedo can increase this ratio by a
factor of $\sim 2-6$.

\subsection{Effects of photospheric albedo and neutral chromosphere}
\label{sec:albedo}
The solar atmosphere above the regions where the X-ray bremsstrahlung
is produced during flares is optically thin, while the dense
photospheric layers below those regions are optically thick to X-ray
photons. As a consequence, photons emitted downwards can be
efficiently Compton backscattered by atomic electrons
in the photosphere
\citep{1972ApJ...171..377T,1973SoPh...29..143S,1978ApJ...219..705B,1995MNRAS.273..837M}.
The observed X-ray spectrum from flares is then a combination of the
primary and albedo photons. The reflectivity is dependent both on the
primary photon spectrum and the flare position on the Sun
relative to the observer, or more precisely, the heliocentric angle of
the source. The backscattered radiation can be significant and modify
the primary X-ray spectrum. More importantly for our analysis here,
when the albedo contribution is taken into account, the number of
electrons required to explain the observed photon flux
changes \citep{2002SoPh..210..407A,2006A&A...446.1157K}.
New fittings taking the albedo correction \citep{2006A&A...446.1157K} into account
give expectedly lower numbers for the electron rate
at footpoints $\dot{N}_{FP} / 10^{35}$ electrons
$s^{-1}$: $18\pm 3$, $69\pm 5$, $0.9\pm 0.1$, $2.1\pm 0.5$, or factors
of 1.8, 2.2, 1.2, 2.5 lower than the electron rate found without
considering the albedo photons. The effect is more pronounced for
flares closer to the centre of the solar disk, while less
significant for limb flares (only 20\% for flare C), as expected. Similar results are found
for the full (spatially integrated) spectrum.

The partially ionized or fully neutral chromosphere is a more
efficient thick-target bremsstrahlung emitter
\cite[e.g.][]{1956PThPh..16..139H,1973SoPh...31..143B,1978ApJ...224..241E},
which can increase HXR flux by a factor of $\sim 3$ in comparison with
fully ionized atmosphere.  The decrease in the ionization of the
target reduces the collisional energy loss of the non-thermal
electrons, thus enhancing the efficiency of the HXR bremsstrahlung.
Applying the thick-target model with a neutral target for the
  footpoint emission, we found the electron rates $\dot{N}_{FP} /
  10^{35}$ electrons $s^{-1}$ for each flare, respectively, $13\pm 2$,
  $60\pm 4$, $0.4\pm 0.1$, $2.0\pm 0.5$, which are a factor of $\sim
  2.5$ smaller than the values found for a fully ionized target.
Similar results are found for the spatially integrated spectrum.

Considering the thick-target model with both the albedo component and
a neutral target, the values for $\dot{N}_{FP}$ for each flare are
$6\pm 1$, $27\pm 2$, $0.4\pm 0.1$, $0.8\pm 0.2$ in units of $10^{35}$
electrons $s^{-1}$. With $\dot{N}_{LT}$ found in the previous
section, the values for the ratio $\dot{N}_{LT}/\dot{N}_{FP}$ are $9.3
\pm 3.0$, $10.1 \pm 2.3$, $4.5\pm 1.4$, $7.8 \pm 5.2$, with an average
$7.9$.

\section{Discussion}
\label{sec:discussion}

The value $\dot{N}_{LT}/\dot{N}_{FP}>1$ found for the flares suggests
either (i) the emissivity per unit of length in coronal sources that
is more efficient than free propagation or (ii) additional energy
losses in the footpoint region. These electron rates also imply that
the main acceleration/injection region is near the looptop, at least
for these flares. We would possibly expect to find
$\dot{N}_{LT}/\dot{N}_{FP} \le 1$ if the acceleration region is not
near the looptop or even if the electrons are accelerated along the
loop, due either to stochastic electric fields
\citep[e.~g.][]{2004ApJ...608..540V,2006A&A...449..749T,2008ApJ...687L.111B,2010ApJ...720.1603G}
or to re-acceleration by beam-generated Langmuir waves
\citep{2012A&A...539A..43K}. We now discuss the result of $
\dot{N}_{LT}/\dot{N}_{FP} >1$ in terms of magnetic trapping, injection
properties, and pitch-angle scattering.

\subsection{Magnetic trapping}
The simplest scenario which can be involved to explain the excess of electrons in the loot-top area
is magnetic trapping due to adiabatic mirroring in a converging magnetic field.
The convergence of the magnetic field, characterized by its mirror ratio
$\sigma =B_{FP}/B_{LT}$ and losscone angle $\alpha_0=\mathrm{acos}(\mu_0)=\mathrm{asin}(\sigma^{-1/2})$,
and the conservation of the magnetic moment of the fast electrons divide the injected
particle population $I$ in a trapped fraction $T$, where the pitch-angle $\alpha>\alpha_0$,
while the other precipitating fraction $P$ (with $\alpha < \alpha_0$) escapes the trap
and precipitates into the dense chromosphere.  Defining $\xi$ as the fraction
of the injected population that remains trapped, we have
\begin{eqnarray}
  I&=&T+P, \\
  T&=&\xi I, \\
  P&=&(1-\xi) I,
\end{eqnarray}
relating this scenario to our results:
\begin{eqnarray}
  \dot{N}_{FP} &=& P, \\
  \dot{N}_{LT} &=& T+P,
\end{eqnarray}
\begin{equation}
  \frac{\dot{N}_{LT}}{ \dot{N}_{FP}}=\frac{1}{1-\xi},
\end{equation}
where
\begin{eqnarray}
  \xi &=& 1-\frac{\dot{N}_{FP}}{ \dot{N}_{LT}},
  \label{eq:traprate}
\end{eqnarray}
where the footpoint electron rate $\dot{N}_{FP}$ indicates the
precipitated fraction $P$, while $\dot{N}_{LT}$ indicates the
  fraction of electrons that pass through the looptop, i.e. both the
  trapped fraction and the electrons that precipitate directly.

Let us consider two typical pitch-angle distributions of energetic
electrons: (i) isotropic $f(\mu)=\mathrm{constant}$ and (ii) beamed
$f(\mu)$, defined as a Gaussian distribution centred along the loop
with spread $\Delta\mu=0.1$. For these two models, one finds the following
relations
\begin{equation}
  \frac{T}{I} =\xi = \frac{\int_{-\mu_0}^{\mu_0}
    f(\mu)d\mu}{\int_{-1}^{1} f(\mu)d\mu}
 \label{eq:int_tr1}
\end{equation}

\begin{equation}
  \frac{P}{I} = 1-\xi = \frac{\int_{-1}^{-\mu_0} f(\mu)d\mu +
    \int_{\mu_0}^{1} f(\mu)d\mu}{\int_{-1}^{1} f(\mu)d\mu}
  \label{eq:int_tr2}
\end{equation}
Eq. \ref{eq:int_tr1} and \ref{eq:int_tr2} can be solved for the
losscone value $\mu_0$, and hence $\sigma $.  For the isotropic case,
Eq. \ref{eq:int_tr1} and \ref{eq:int_tr2} reduce to
$T/I=\mu_0=\xi$ and $P/I=(1-\mu_0)=(1-\xi)$.  Solving these equations
using the observed values of $\dot{N}_{LT}$ and $\dot{N}_{FP}$, we find
the values for $\xi$ and $\sigma $, which are presented in Table
\ref{tab:mirror}. Here, we stress that this simple model
  is meant as a way to divide the injected population in terms of the
  losscone, representing an {\em upper} limit for the derived mirror ratio
  values, since the accumulation of electrons is not considered in the model.

Under the assumption of an isotropic injection, $\sigma$ values are
small ($\sigma \gtrsim 1$) and in agreement with values found by
\citet{1999ApJ...517..977A} and \citet{2007A&A...461..315T}. In the
case of the narrow beam, we found that $\sigma \gtrsim 5$, which is
higher than the range of mirror ratios found by
\cite{2005A&A...436..347C} for microwave data, without assuming a
particular pitch-angle distribution, although their determination is
biased by a higher weight of the loop-leg source and offers only a lower
limit of the mirror ratio. When considering our results for the case
with albedo and neutral target, we found $2.5 \le \sigma \le 5$ for
the isotropic case and $\sigma \gtrsim 17-42$, for the beamed
injection case. As mentioned above, the values for the
required mirror ratio are just upper bounds. The observations could
be explained by lower mirror ratios and sustained injection which cause
accumulation of electrons at the looptop.

\begin{table}[!h]
  \caption{Estimated values for trapping fraction $\xi$ and mirror
    ratio $\sigma$.}
  \label{tab:mirror}
  \centering
  \begin{tabular}{lcccc}
    \hline\hline
    Flare &   A & B & C & D\\
    \hline
    ionized target, no albedo & & & & \\
    trapped fraction $\xi$ & 0.4 &0.4 &0.3 &0.3\\
    $\sigma $ isotropic &  1.2 &1.2 &1.1 &1.1\\
    $\sigma $ beam      & 6 & 6 & 5& 5\\
    \hline
    neutral target, albedo & & & & \\
    trapped fraction $\xi$ & 0.9 &0.9 &0.8 &0.9\\
    $\sigma $ isotropic &  5 &5 &2.5 &4\\
    $\sigma $ beam & 36 & 42 & 17& 32\\
    \hline
\end{tabular}
\end{table}

\subsection{Pitch-angle scattering}

The accelerated electrons in the flaring loop are usually assumed to
be trapped within an adiabatic magnetic trap. Two approaches can
be used to describe their dynamics and spectra, namely Coulomb
collisions \citep{1976MNRAS.176...15M,1985SoPh...99..231M} and
interactions between electrons and waves in the plasma
\citep{1976ApJ...208..595W,1987SoPh..114..127B,2002SoPh..211..135S}.
Observed energy-dependent time delays of the 20-200 keV HXR emission
favour the Coulomb collision hypothesis
\citep{2005psci.book.....A}. The time delays of $1-10$ s for smoothed
components of the emission follow the law $\epsilon^{3/2}$ ($\epsilon$
is photon energy) and can be explained by Coulomb scattering of
non-thermal electrons by dense background plasma
\citep{1997ApJ...487..936A}. However, these observed time delays can
also be explained by a magnetic trap plus pitch-angle
  scattering by plasma turbulence \citep{1999ARep...43..838S}.

\citet{1987SoPh..114..127B}, following \cite{1966JGR....71....1K},
describe three diffusion regimes of pitch-angle diffusion: {\it weak
  diffusion regime}, {\it moderate diffusion regime}, and {\it
  strong diffusion regime}. The three regimes of pitch-angle diffusion
are characterised by the mean time of pitch-angle scattering $t_d$,
flight time through the loop $t_c = L/v$, where $v$ is the particle
speed, and mirror ratio $\sigma = B_{FP}/B_{LT}$: weak ($t_d > \sigma
t_c$), intermediate, or strong in \cite{1966JGR....71....1K}
description, ($t_c < t_d < \sigma t_c$), and strong ($t_d < t_c$). In
terms of pitch-angle distribution, the weak regime generally means that the
electrons are weakly (Coulomb) scattered in a timescale
of the bounce period and that the losscone is empty. In the
intermediate regime, the electrons can travel through the loop and are
scattered into the losscone, which can be considered full, and the
pitch-angle distribution is nearly isotropic. The strong regime means
that a electron can change its direction many times during one bounce
period. This regime can only take place by wave-particle interactions,
as a strong collisional diffusion also means that the fast electrons
will quickly lose their energy and thermalize.

To explain the observations in a scenario without magnetic trapping,
i.e. no coronal convergence of the magnetic field, the pitch angle
scattering should be strong $t_d/t_c \sim
(\dot{N}_{LT}/\dot{N}_{FP})^{-1}$, with the values of $0.1 < t_d/t_c <
0.7$. Let us consider that the HXR-producing electrons are
magnetically trapped only at the looptop where the emission appears, i.e.
the mirror points are high in the coronal field. In this case, it is
reasonable to expect that $\sigma$ would not be much higher than
$~2$. This scenario can be justified because there is no emission
detected along the loop legs, which would be expected if the electrons
were to bounce between the mirrorpoints near the footpoints. From our
results of the trapping analysis in Table \ref{tab:mirror}, we find
$\sigma > 2$, except when an isotropic injection and an ionized target
are considered. Assuming that at the looptop $1 < \sigma < 2$ but a higher
value of $\sigma$ is required to trap the electrons, efficient
scattering is needed at the looptop to change the pitch-angle distribution
and allow the electrons to be magnetically trapped. Also, the
isotropic injection implies strong scattering in the acceleration
process. Although the electron accumulation at the looptop can be
explained by a trap-plus-precipitation model with modest mirror ratios
$\sigma<2$, we concluded that the pitch-angle distribution must be
wide enough to be divided by the losscone, or else it will result in
either a fully trapped (as proposed by \citet{2002ApJ...580L.185M} due
to an anisotropic pitch-angle distribution orthogonal do the magnetic
field, the so-called pancake distribution) or fully precipitated
scenario (as in the classical thick-target model). This requirement
can be fulfilled by moderate/strong pitch-angle scattering processes.
The presence of strong pitch-angle scattering is particularly
important for stochastic acceleration models, where the transport of
electrons strongly influences the acceleration rate of particles
\citep[see][as a recent review]{2012ApJ...754..103B}.

\subsection{Instrumental effects}

\begin{figure*}
\resizebox{\hsize}{!}{\includegraphics[angle=0]{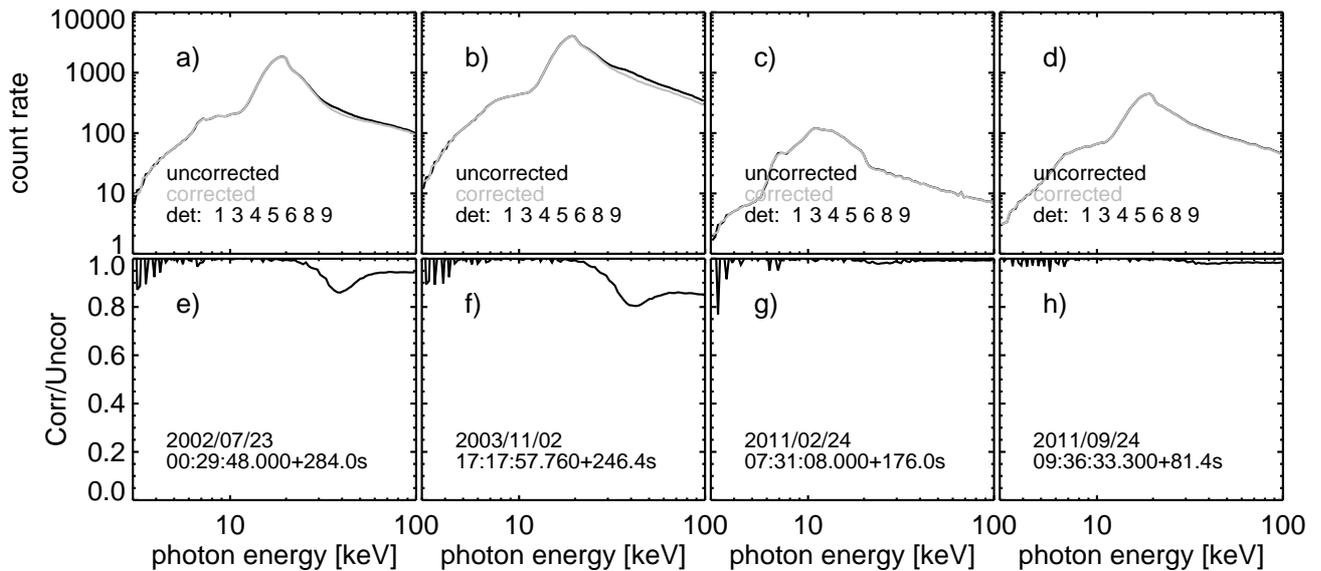}}
\caption{Pileup effect in the count spectra for flares A, B, C, and D
  (time intervals in Table \ref{tab:eventinfo} and noted in each
  frame). {\em Top row:} corrected (gray) and uncorrected (black)
  count spectra. {\em Bottom row:} the ratio of pileup corrected and
  uncorrected count spectra for each flare.}
\label{fig:pileup}
\end{figure*}

Because the events we selected for this analysis are GOES M and X
class flares, it is important to check for pileup
\citep{2002SoPh..210...33S}. This effect, depending on attenuator
state, is usually stronger in the range 20-50 keV, which is the range
where the looptop sources were identified. We tested the importance
of pileup, comparing the ratio of the corrected and uncorrected count
rate spectra of each flare. These spectra and their ratio are
presented in Fig. \ref{fig:pileup}, taken for the time intervals
presented in Table \ref{tab:eventinfo}. Presently, it is not possible
to correct for pileup in imaging spectroscopy; however, we can assess
its effects in the spatially integrated spectrum and estimate the
errors in the looptop and footpoint spectra. Taking the minimum of
the ratio between corrected and uncorrected count rate spectra as the
maximum error due to pileup, we find the values of 14\% and 20\% (both
around 40 keV) for flares 2002 July 23 and 2003 November 02, while
errors due to pileup are lower than 2\% for flares 2011 February 24
and 2011 September 24. These errors are much smaller than the errors
from the fittings propagated to the final ratio
$\dot{N}_{LT}/\dot{N}_{FP}$: 35\%, 24\%, 27\%, and 64\%. Consequently,
pileup effects do not seem to be significant in our findings.

\section{Summary}
\label{sec:conclusion}

We presented an analysis of the electron distribution in footpoints
and coronal sources for four well-observed solar flares and, for the
first time, found the electron rate required to explain the coronal
and chromospheric X-ray emissions. Using imaging spectroscopy
techniques, the X-ray spectra of the coronal and footpoint sources
are obtained and fitted with thin- and thick-target models,
respectively. The thick-target model directly provides the electron
rate $\dot{N}_{FP}$ necessary to explain the observed footpoint
emission. To obtain the electron rate $\dot{N}_{LT}$ at the looptop
source from the thin-target model, it is also necessary to find the
length of the source $L$ and the thermal plasma density $n$
(Eq. \ref{eq:nlt}). The length $L$ was measured for each flare in
CLEAN maps in an energy range where the coronal source was resolved at
50\% of the maximum of the image and thermal contribution can be
safely neglected, while the plasma density $n$ was deduced from the
emission measure $EM$ (from spectral fittings) and source volume
estimated from $\sim 10$ keV source size.

Assuming no albedo and a fully ionized target, we found an average
\ratio ~ of $1.6$.  The ratio \ratio ~ is further enhanced when
considering the Compton back-scattered photons, i.e. albedo
\citep[e.g.][]{2006A&A...446.1157K}, and a neutral target in the
chromosphere \citep[e.g.][]{1973SoPh...28..151B}.  Both processes
reduce the required electron rate to match the observed HXR photon
flux, effectively reducing $\dot{N}_{FP}$. Under these considerations,
the average ratio found is $\sim 8$. The ratio \ratio$>1$ could
suggest that the acceleration/injection region is near the looptop,
at least for these flares. In previous studies using microwave
emission maps, homogeneous electron distribution along the loop
\citep{2001ApJ...557..880K} or high accumulation of electrons at the
looptop \citep{2002ApJ...580L.185M} were proposed to explain the
loop-like morphology of the 17 and 34 GHz
emission. \citet{2002ApJ...580L.185M} concluded that a ratio of
electrons at the looptop to footpoint in the range $\sim 10-100$ is
required to reproduce the observations of the four flares considered
in their study. We note, however, that those results only account for
the trapped population, without considering the number of precipitated
electrons required to produce HXR at chromospheric footpoints. In
fact, their solution requires a pancake pitch-angle distribution at
the looptop, which would produce very weak HXR chromospheric emission
(if any at all) due to the lack of direct precipitation. Moreover,
there is extensive evidence of microwave sources associated with HXR
footpoint emission
\citep[e.g.][]{1995ApJ...454..522K,1997ApJ...489..976N,2000ApJ...531.1109L,2006IAUS..233..334K,2006ApJ...643.1271S,2009SoPh..260..135K},
indicating that the trapping conditions vary strongly from flare to
flare, as one might expect. In fact, evidence for how trapping
conditions are different for each flare is presented by
\citet{1994R&QE...37..557M}, showing that the ratio between the number
of trapped high-energy electrons (i.e. produce microwave emission) and
precipitated low-energy electrons (i.e. electrons that produce HXR
emission) can vary by three orders of magnitude, depending
on the electron lifetime value inside the trap and duration of the
injection. We note that all those works compared two different
electron energy ranges, while in this paper our results are presented
for electrons in the deka-keV energy range.

We considered a collisionless magnetic trapping to explain the
electron rates found (for fully ionized target and no albedo) and
found that the required mirror ratios are in the range $1.1 \lesssim
\sigma \lesssim 6$, while for a neutral chromospheric target and
accounting for albedo photons, $2.5 \lesssim \sigma \lesssim
17-42$. These values reflect an upper boundary for the mirror ratio,
as our simple model does not account for electron accumulation over a
sustained injection. Therefore, the observations can possibly be
explained considering a trap-plus-precipitation scenario with modest
mirror ratio values. However, we note that the pitch-angle
distribution must be wide enough to be divided into trapped and
precipitated fractions by the trap losscone. This requirement
indicates that moderate/strong pitch-angle scattering must take place
in the acceleration site or during the transport along the loop.

If we consider a scenario without significative convergence of the
coronal field towards the chromosphere, i.e. no efficient magnetic
trapping, the pitch-angle scattering time required to explain the
observations appears to be in the range $0.1-0.7$ of the time required
for an electron to cross the length of the loop. The observations do
not support very strong scattering with values smaller than $0.1$,
which is sometimes required for efficient stochastic
acceleration. This conclusion indicates that the observed coronal
source is not the acceleration region, which could be in a smaller or
larger portion of the loop. Hence the coronal HXR source is mainly due
to transport processes. Equally, it can also mean that the stochastic
acceleration models with strong pitch-angle scattering are not the
dominant mechanism for electron acceleration at these energies.
Conclusions in favour of the trap-plus-precipitation model are given
by other studies
\citep[e.g.][]{1999ApJ...517..977A,2007A&A...461..315T}.
Nevertheless, the effect of moderate/strong scattering can possibly
explain the formation of the HXR looptop source or at least enhance the
effect of the magnetic trapping. It is our impression that the
scattering cannot be neglected and should be investigated with the
trap-plus-precipitation model for the deka-keV energy range in more
detail.

In conclusion, our results suggest that the accelerated electrons must
be subject to magnetic trapping and/or moderate or strong pitch-angle
scattering, keeping a fraction of the population trapped inside the
coronal loops. This is in full agreement with the
trap-plus-precipitation models first proposed by
\cite{1966PASJ...18...57T} and further developed by many other
authors.

\begin{acknowledgements}
  Financial support by the European Commission through HESPE
  (FP7-SPACE-2010-263086) (PJAS, EPK), the "Radiosun"
  (PEOPLE-2011-IRSES-295272) Networks (EPK), and by the STFC rolling
  grant is gratefully acknowledged.
\end{acknowledgements}
\bibliographystyle{aa}
\bibliography{refs_rhessi}

\end{document}